\documentclass[review,3p,authoryear]{elsarticle}

\usepackage{amssymb}
\usepackage{natbib}
\usepackage{epstopdf}
\usepackage{amscd}
\usepackage{amsmath}
\usepackage{amsfonts}
\usepackage{bm}
\usepackage{bbm}
\usepackage{latexsym}
\usepackage{mathrsfs}
\usepackage{amsthm}
\usepackage{mathrsfs}
\usepackage{booktabs}
\usepackage{caption}
\usepackage{rotating}
\usepackage{multirow}
\usepackage{lscape}
\usepackage{subfig}
\usepackage{color}

\def\sc{\mathsf{c}}

\def\bdelta{\mbox{\boldmath $\delta$}}
\def\bDelta{\mbox{\boldmath $\Delta$}}

\def\bGamma{\mbox{\boldmath $\Gamma$}}

\def\bxi{\mbox{\boldmath $\xi$}}

\def\blambda{\mbox{\boldmath $\lambda$}}
\def\btheta{\mbox{\boldmath $\theta$}}

\def\bXi{\mathbf{\Xi}} 

\def\br{\mathbf{r}}
\def\bu{\mathbf{u}}

\def\by{\mathbf{y}}

\def\0{\mbox{\bf{0}}}

\def\bC{\mathbf{C}}

\def\bN{\mathbf{N}}
\def\bR{\mathbf{R}}

\def\bX{\mathbf{X}}

\def\bA{\mathbf{A}}

\def\bK{\mathbf{K}}
\def\bQ{\mathbf{Q}}

\def\bB{\mathbf{B}}
\def\bD{\mathbf{D}}

\newcommand\bbone{\ensuremath{\mathbbm{1}}}
\newcommand{\suchthat}{\;\ifnum\currentgrouptype=16 \middle\fi|\;}
\def\qmo{``}
\def\qmc{''}
\def\qmcsp{'' } 

\begin{document}

\begin{frontmatter}



\title{\LARGE Portfolio Optimisation Under Flexible\\ Dynamic Dependence Modelling}

\author[bernardi]{Mauro Bernardi\corref{corresp}}
\author[catania]{Leopoldo Catania}
\ead[catania]{leopoldo.catania@uniroma2.it}
\address[bernardi]{Department of Statistical Sciences, University of Padova, Padova, Italy.}
\address[catania]{Department of Economics and Finance, University of Rome, Tor Vergata, Rome, Italy}
\cortext[corresp]{Department of Statistical Sciences, University of Padova, Via C. Battisti, 241/243, 35121, Padova, Italy. \texttt{mauro.bernardi@unipd.it}, tel.; +39.049.82.74.165, web page: \texttt{http://homes.stat.unipd.it/maurobernardi/}.}

\begin{abstract}
Signals coming from multivariate higher order conditional moments as well as the information contained in exogenous covariates, can be effectively exploited by rational investors to allocate their wealth among different risky investment opportunities. This paper proposes a new flexible dynamic copula model being able to explain and forecast the time--varying shape of large dimensional asset returns distributions. 
Moreover, we let the univariate marginal distributions to be driven by an updating mechanism based on the scaled score of the conditional distribution. This framework allows us to introduce time--variation in up to the fourth moment of the conditional distribution. The time--varying dependence pattern is subsequently modelled as function of a latent Markov Switching process, allowing also for the inclusion of exogenous covariates in the dynamic updating equation. We empirically assess that the proposed model substantially improves the optimal portfolio allocation of rational investors maximising their expected utility.
%
%
\end{abstract}
\end{frontmatter}
%
%
\section{Introduction}
\label{sec:intro}
\noindent The problem of wealth allocation is a key driver for past and present economic studies. Since the seminal contribution of \cite{markowitz.1952}, both the industry and the researchers, have been highly involved in finding the optimal way to improve the performance of allocation strategies. During last decades, the Markowitz's approach has been widely criticised for the assumptions behind its simple mean--variance framework. In particular, the most critical points regard the homoskedasticity assumption and, in general, the Gaussian assumption for the joint returns distribution.
Recently, researchers and practitioners found that the departure from the simple Markowitz's mean--variance approach, considering, for example, higher moments, can provide substantial improvements of the asset allocation strategies in terms of cumulative returns, see, e.g., \cite{jondeau_rockinger.2003}, \cite{lai_etal.2006}, \cite{harvey_etal.2010} and \cite{holly_etal.2011}. Furthermore, as for the conditional second moment, there have also been several attempts to model the conditional skewness and kurtosis of financial time series.
Indeed, starting from the seminal contribution of \cite{hansen.1994}, in the last two decades, the financial econometric literature has been focusing on modelling time--varying higher--order moments. One of the main issues  is to figure out the correct way to introduce time--variation in skewness and kurtosis. Most important contributions in this sense are those of \cite{hansen.1994}, \cite{theodossiou_1998}, \cite{harvey_siddique.1999}, \cite{jondeau_rockinger.2003}, \cite{deluca_etal.2006} and \cite{jondeau_rockinger.2012}, who extend the model of \cite{hansen.1994} and the class of ARCH models of \cite{engle.1982} and \cite{bollerslev.1986}. However, when dealing with higher--order conditional moments' dynamics in a multivariate context, the complexity increases principally for the high number of parameters needed and the major concern involves the overall tractability of the resulting model. As regards the co--skewness, main results are referred to \cite{deluca_etal.2006}, \cite{deluca_loperfido.2012} and \cite{franceschini_loperfido.2010}, while for the co--kurtosis, up to our knowledge, there are not any contributions that impose a conditional dynamics on the fourth moment.\newline
\indent The problem of dealing with higher order co--moments is strictly related to the general issue of describing the dependence structure affecting the multivariate financial time series. Since the seminal contribution of \cite{sklar.1959}, copula models have been effectively used as an alternative flexible tool to the classical multivariate Gaussian models. Indeed, in the last years, copulas theory has played a central role for the financial econometric literature, as widely documented by \cite{mcneil_etal.2005} and \cite{joe.2014}. A comprehensive treatments of recent developments in this field can be found in the recent book of \cite{durante_sempi.2015}. Furthermore, starting from the seminal contribution of \cite{patton.2006a} and \cite{jondeau_rockinger.2006b}, copula theory has been further extended to cope with dynamic evolving dependence structures. Time--varying copulas have been successfully employed in empirical works such as in \cite{rodriguez.2007}, \cite{delirasalvatierra_patton.2015} and \cite{bernardi_catania.2015a}.\newline
\indent An effective way to improve the models in sample fit as well as their forecast ability, is to include exogenous information. Nevertheless, most econometric models that aim to explain the joint dependence between economic factors, only consider endogenous information. By the way, it is natural to believe that, the dynamics affecting the relations between industries, and more generally between economies, are highly affected by exogenous, as well as, endogenous factors. It is worth noting that, the economic and financial literature focusing on univariate models, makes extensive use of exogenous information, see, e.g., \cite{pagan.1971} and \cite{baillie_1980} in the ARMAX context and \cite{hwang_satchell.2005} for the GARCHX volatility model. Their empirical results confirm the superior ability of the models exploiting exogenous information, as compare with more simple alternatives. However, despite the documented usefulness of accounting for exogenous information in univariate time series models, in the financial econometric literature that focuses on the dynamic dependence modelling, there are only few examples of multivariate models dealing with exogenous information, see, e.g., \cite{delirasalvatierra_patton.2015}. This lack is principally related to the difficulty of effectively including such information in such a way to retain model tractability, especially when the dimension of the observed variables is greater than 2. However, the contribution of including exogenous information in order to describe the dependence structure of economies or assets cannot be neglected. Let think about how the recent Global Financial Crises (GFC) of 2007--2008 influenced the dependence between economies. It is undoubtedly true that such a dramatic event drastically changed the choices of wealth allocation of economic agents. Indeed, capitals have moved away from the financial sector, and more generally from speculative instruments, such as options and futures, opening positions on less riskier assets such as short term government bonds or liquidity funds. In this reallocation of resources, it is obvious that, the behaviour of the sovereign interest rates dynamics, for examples, plays a central role in determining how much of the total wealth is reallocated, and hence how the dependence structure of the remaining assets is affected.\newline
%
%
%
\indent In this paper, we propose a new Markov switching dynamic copula model being also able to deal with exogenous information in a fully multivariate setting using elliptical copulas. More precisely, the tractability induced by the copula methodology in dealing with marginal and joint separability is gathered with a new dynamic conditional dependence structure that relies on the presence of exogenous covariates and the inclusion of a latent Markovian structure, in order to enhance the model ability to account for time--varying higher moments. In this way, we are able to incorporate relevant information such as third and fourth conditional moments, as well as economic factors, in a portfolio optimisation framework. Throughout the paper, we will distinguish the marginal modelisation from the dependence structure linking individual asset into a multivariate framework. This distinction is available thanks to the use of copulas, and more precisely, to the \cite{sklar.1959} Theorem and to the Inference Function for Margins (IFM) technique of \cite{godambe_1960} and \cite{mcleish_small.1988}.
As regards the univariate specification, we develop a new flexible parametric model accounting for time--varying conditional moments up to the fourth of the univariate distribution. In particular, we rely on the Generalised Autoregressive Score (GAS), also known as Dynamic Conditional Score (DCS), framework recently proposed by \cite{creal_etal.2013} and \cite{harvey.2013} in order to update the dynamics of time--varying parameters using the scaled score of the conditional distribution. This specification allows us to take account of a variety of well know stylised facts affecting the financial time series, like time varying moments, skewness and excess of kurtosis, as widely discussed by \cite{mcneil_etal.2005}. Particularly interesting is the ability of the employed univariate model to approximate the unknown dynamics of the conditional density of returns by means of a filter based on the scaled score of the conditional distribution. This approach has been recently proved to be consistent with the goal of describing time variation for densities' parameters, see \citep{koopman_etal.2015}.
%
\indent As stated before, the multivariate modelisation is carried on by using a copula function that links the dependence structure of each univariate series. This solution has been proved to provide highly flexible and reliable models for multivariate financial time series analysis, see e.g. \cite{patton.2006b}, \cite{joe.2014}, and the references quoted therein. Moreover, the seminal work of \cite{patton.2006a}, also extend the original IFM two step estimator to the time--varying copula parameters framework. This fundamental result allows to effectively model the time--varying behaviour of linear as well as non--linear dependence patterns of financial returns. As discussed by \cite{mcneil_etal.2005}, the flexibility in modelling linear and non--linear dependence patterns is one of the main stylised fact affecting multivariate financial time series. Indeed, as a direct consequence, econometric models dealing with multivariate financial time series, are required to account for this aspect, see, e.g., \cite{delirasalvatierra_patton.2015}. For this reason, we propose a new specification for the evolution of the conditional correlation matrix of elliptical copulas relying on the famous Dynamic Conditional Correlation (DCC) specification of \cite{engle.2002} and \cite{tse_tsui.2002}, and the extension of \cite{cappiello_etal.2006}. Moreover, as in \cite{billio_caporin.2005} and, more recently, \cite{bernardi_catania.2015a}, we also allow for the dynamic evolution of the conditional correlation matrix to depend upon the realisation of a first order Markovian process. More precisely, our model allows for different dynamic behaviours of the underlying dependence process in each specific state of the nature, as in \cite{billio_caporin.2005} and \cite{bernardi_catania.2015a}, but it additionally includes exogenous regressors in the dependence structure conditional to the latent state. The proposed model can effectively describe tranquil as well as turbulent periods affecting the financial markets, as for example, different phases of the economic or financial cycles. The opportunity to include exogenous covariates into the correlation dynamics is motivated by the observation that, in nowadays financial markets, economic agents usually base their decisions about wealth allocation conditionally on a huge amount of available information coming from multiple sources, which cannot be neglected. Summarising, our new specification is specifically tailored to model time variation in higher order conditional moments and possibly nonlinear time variation in the dependence structure of financial time series including the ability to account for exogenous information.\newline
\indent The empirical part of the paper considers a panel of financial returns in a portfolio optimisation framework using higher order moments of the predictive conditional joint distribution. More precisely, in a similar way as \cite{jondeau_rockinger.2006a}, we consider conditional portfolio moments into the expected utility of a rational investor by means of a Taylor expansion around the future wealth. Our empirical findings support our model since it outperform famous alternatives, like the DCC model. Specifically, the new specification improves the performance of allocation strategies based on the maximisation of the expected utility.\newline
\indent The remaining of the paper is organised as follows. Section \ref{sec:model} introduces the marginal and the joint dynamic model specifications and discusses their main features. Section \ref{sec:asset_allocation} describes the asset allocation problem. Section \ref{sec:empirical_study} deals with the empirical study and Section \ref{sec:conclusion} concludes.
%
\section{The Model}
\label{sec:model}
%
\noindent In this paper, we propose to estimate the parameters governing the copula function and the associated dependence structure separately from those controlling for the dynamic evolution of the marginal distributions. This estimation technique, known as Inference Function for Marginals (IFM), has been proved to be feasible in large parametric spaces and asymptotically consistent for time varying copulas, see, e.g., \cite{patton.2006a}. One of the main advantages of using the IFM estimating procedure concerns its ability to separate the univariate financial returns stylised facts from those regarding the multivariate distribution. This latter aspect, allows the researcher to focus on multivariate dependence structure only in a second moment, when the most appropriate model to describe all the empirical regularities of the observed univariate series has been selected and estimated. Whenever the marginal models are able to capture all the univariate empirical regularities of the marginal series, the only interestingly phenomenon to study is the remaining dependence structure. It follows that, the choice of an appropriate model that acts as a filter for the univariate series plays an important role for the subsequent multivariate analysis, see, e.g., \cite{mcneil_etal.2005}.\newline
\subsection{Marignal models}
\label{sec:marginal_model}
\noindent The financial econometrics literature offers several valid alternatives to model the conditional distribution of financial returns. In a fully parametric setting, following the same arguments presented by \cite{cox.1981}, the first issue is to choose between the classes of observation driven and parameter driven models. The former can be well represented by the GARCH family of \cite{engle.1982} and \cite{bollerslev.1986}, while the latter is typically associated with the class of state space models; see for example \cite{harvey_proietti.2005} and \cite{durbin_koopman.2012}. However, the Generalised Autoregressive Score (GAS) framework, recently introduced by \cite{creal_etal.2013} and \cite{harvey.2013} is gaining lots of consideration by econometricians in many fields of time series analysis. Under the \cite{cox.1981} classification, the GAS models can be considered as a class of observation driven models, with the usual consequences of having a closed form expression for the likelihood and ease of evaluation. However, as noted by \cite{koopman_etal.2015}, the class of GAS models is similar to the class of state space models in the sense that both approaches make use of the information coming from the entire conditional distribution instead of using only that coming from the conditional expected value, as for example in the GARCH framework.
Indeed, the key feature of GAS models is that the score of the conditional density is used as the forcing variable into the updating equation of a time--varying parameter. In the recent financial and econometrics literature, many reasons have been argued to support the adoption of score--based rules as a general updating mechanism of time--varying parameters, see, e.g., \cite{harvey.2013} and \cite{creal_etal.2013}, just to quote a few of them. Moreover, the flexibility of the GAS framework makes this class of models nested with a huge amount of famous econometrics models such as, for example, some of the ARCH--type models of \cite{engle.1982} and \cite{bollerslev.1986} for volatility modelling, and also the MEM, ACD and ACI models of \cite{engle.2002b}, \cite{engle_gallo.2006}, \cite{engle_russell.1998} and \cite{russell.1999}, respectively. Finally, one of the practical implications of using this framework in order to update the time--varying parameters, is that it avoids the problem of using a non--adequate forcing variable when the choice of it is not so obvious as for dynamic copula models; see, e.g., \cite{bernardi_catania.2015a}.\newline
%
\indent Formally, we assume that the $i$--th return at time $t$, for $t=1,2,\dots,T$, is conditionally distributed according to the filtration $\mathcal{F}_{i,t-1}=\sigma\left(y_{i,1},y_{i,2},\dots,y_{i,t-1}\right)$ as
\begin{equation}
y_{i,t}\sim\mathcal{AST}\left(y_{i,t};\mu_{i,t},\sigma_{i,t},\gamma_{i,t},\nu_{i,t}\mid\mathcal{F}_{t-1}\right),
\end{equation}
where $\mathcal{AST}\left(y_{i,t};\cdot\right)$ denotes the Asymmetric Student--t distribution with equal left and right tail decay of \cite{zhu_galbraith.2010} with time varying location $\mu_{i,t}\in\mathbb{R}$, scale $\sigma_{i,t}\in\mathbb{R}^+$, shape $\nu_{i,t}\in\left(4,\infty\right)$ and skewness $\gamma_{i,t}\in\left(0,1\right)$ parameters and density given by
\begin{equation}
f_\mathcal{AST}\left(y_{i,t};\mu_{i,t},\sigma_{i,t},\gamma_{i,t},\nu_{i,t}\right)=\begin{cases}
\frac{1}{\sigma_{i,t}}\left[1+\frac{1}{\nu_{i,t}}\left(\frac{y_{i,t}-\mu_{i,t}}{2\gamma_{i,t}\sigma_{i,t} K\left(\nu_{i,t}\right)}\right)\right]^{-\frac{\left(\nu_{i,t}+1\right)}{2}},\quad& y_{i,t}\le\mu_{i,t}\nonumber\\
\frac{1}{\sigma_{i,t}}\left[1+\frac{1}{\nu_{i,t}}\left(\frac{y_{i,t}-\mu_{i,t}}{2(1-\gamma_{i,t})\sigma_{i,t} K\left(\nu_{i,t}\right)}\right)\right]^{-\frac{\left(\nu_{i,t}+1\right)}{2}},\quad& y_{i,t}>\mu_{i,t},\nonumber\\
\end{cases}
\end{equation}
where $K\left(x\right) = \Gamma\left(\left(x+1\right)/2\right)/\left[\sqrt{\pi x}\Gamma\left(x/2\right)\right]$ and $\Gamma\left(\cdot\right)$ is the gamma function. We also define $\btheta_{i,t} = \left(\mu_{i,t},\sigma_{i,t},\gamma_{i,t},\nu_{i,t}\right)^\prime\in\mathbb{R}\times\mathbb{R}^+\times\left(0,1\right)\times\left(4,\infty\right)$ to be a vector containing all the time--varying parameters. Furthermore, let $h:\mathbb{R}^4\to\mathbb{R}\times\mathbb{R}^+\times\left(0,1\right)\times\left(4,\infty\right)$ be a continuous, $\mathcal{F}_{t-1}$ measurable, twice differentiable, vector--valued mapping function, such that $\btheta_t = h\left(\tilde{\btheta}_{i,t}\right)$, where $\tilde\btheta_{i,t}\in\mathbb{R}^4$ is a proper reparametrisation of $\btheta_{i,t}$, for all $t=1,2,\dots,T$. In our context, a convenient choice for $h\left(\cdot\right)$ is
\begin{equation}
h\left(\tilde{\btheta}_{i,t}\right) = \begin{cases}
\mu_{i,t}=\mu_{i,t}\\
\sigma_{i,t}=\exp\left(\tilde\sigma_{i,t}\right)\\
\gamma_{i,t}=\left[1 + \exp\left(-\tilde\gamma_{i,t}\right)\right]^{-1}\\
\nu_{i,t}=4+\exp\left(\tilde\nu_{i,t}\right).
\end{cases}
\end{equation}
We let the time--varying reparameterised vector $\tilde\btheta_{i,t}=\left(\mu_{i,t},\tilde\sigma_{i,t},\tilde\gamma_{i,t},\tilde\nu_{i,t}\right)^\prime$ to be updated using the score of the conditional distribution of $y_{i,t}$, exploiting the GAS dynamic where the Fisher information matrix of the conditional distribution is used as scaling matrix, as suggested by \cite{creal_etal.2013} and \cite{harvey.2013}. Specifically, we consider the following updating mechanism for the parameter dynnamics
\begin{align}
 \begin{pmatrix}
 \mu_{i,t+1}\\
  \tilde\sigma_{i,t+1}\\
  \tilde\gamma_{i,t+1}\\
  \tilde\nu_{i,t+1}
 \end{pmatrix} =
  \begin{pmatrix}
 \omega_{\mu_i}\\
 \omega_{\sigma_i}\\
 \omega_{\gamma_i}\\
 \omega_{\nu_i}
 \end{pmatrix} +
   \begin{pmatrix}
 \alpha_{\mu_i} & 0 & 0 & 0\\
  0 & \alpha_{\sigma_i} & 0 & 0\\
  0 & 0 & \alpha_{\gamma_i} & 0 \\
  0 & 0 & 0 & \alpha_{\nu_i}
 \end{pmatrix} \tilde{s}_{i,t}  +
    \begin{pmatrix}
 \beta_{\mu_i} & 0 & 0 & 0\\
  0 & \beta_{\sigma_i} & 0 & 0\\
  0 & 0 & \beta_{\gamma_i} & 0 \\
  0 & 0 & 0 & \beta_{\nu_i}
 \end{pmatrix}
 \begin{pmatrix}
 \mu_{i,t}\\
  \tilde\sigma_{i,t}\\
  \tilde\gamma_{i,t}\\
  \tilde\nu_{i,t}
 \end{pmatrix},
 \label{eq:gas_empDyn}
\end{align}
where $\tilde s_t = \mathcal{J}\left(\tilde\btheta_{i,t}\right)^{-1}\mathcal{I}\left(\btheta_{i,t}\right)^{-1}\nabla\left(y_{i,t};\btheta_{i,t}\right)$ is the scaled score of the reparameterised conditional distribution of $y_{i,t}$. The quantity $\tilde s_t$ depends on the product of three matrices defined as
\begin{align}
\mathcal{J}\left(\tilde\btheta_{i,t}\right) &= \frac{\partial\mathrm{h}\left(\tilde{\boldsymbol{\theta}}_{i,t}\right)}{\partial\tilde\btheta_{i,t}}\\
\mathcal{I}\left(\btheta_{i,t}\right) & = \mathbb{E}\left[\nabla\left(y_{i,t};\btheta_{i,t}\right)\times\nabla\left(y_{i,t};\btheta_{i,t}\right)^\prime\mid\mathcal{F}_{t-1}\right]\\
\nabla\left(y_{i,t};\btheta_{i,t}\right) &= \frac{\partial\ln f_\mathcal{AST}\left(y_{i,t};\btheta_{i,t}\right)}{\partial{\btheta_{i,t}}},
\end{align}
where $\mathcal{J}\left(\tilde\btheta_{i,t}\right)$ is the Jacobian of the mapping function $h\left(\cdot\right)$, i.e.
\begin{equation}
  \mathcal{J}\left(\tilde\btheta_{i,t}\right) = \begin{pmatrix}
                                                  1 & 0 & 0 & 0 \\
                                                  0 & \exp\{\tilde\sigma_{i,t}\} & 0 & 0 \\
                                                  0 & 0 & -\exp\{-\tilde\gamma_{i,t}\}\left(1+\exp\{-\tilde\gamma_{i,t}\}\right)^{-2} & 0 \\
                                                  0 & 0 & 0 & \exp\{\tilde\nu_{i,t}\},
                                                \end{pmatrix},
\end{equation}
where $\mathcal{I}\left(\btheta_{i,t}\right)$ and $\nabla\left(y_{i,t};\btheta_{i,t}\right)$ are the Fisher information matrix and the score of the conditional AST distribution, respectively. For the parameterisation here considered, the Fisher information matrix of the AST distribution is reported in \cite{zhu_galbraith.2010}, while the score is reported in \ref{sec:ScoreAndFisher}.\newline
\indent It is worth noting that, in equation \eqref{eq:gas_empDyn}, a diagonal structure for the matrix of coefficient is considered. Obviously, this is a convenient choice in order to reduce the number of model parameters, but other solutions are possible. In order to guarantee the stationarity of the process defined in equation \eqref{eq:gas_empDyn} we only need to impose the constraint that all the autoregressive coefficients $\beta_k$, $k\in\left(\mu,\sigma,\gamma,\nu\right)$ would be in modulus less then one, since, under a correct model specification, the sequence of scores $\{\tilde s_t, t>0\}$ is a martingale difference.
%
\subsection{Multivariate model}
\label{sec:multivariate_model}
%
\noindent Dynamic time--varying dependence structures of financial assets have been deeply investigated in the recent developments of financial econometrics. One of the most successful and powerful model has been the Dynamic Conditional Correlation (DCC) model of \cite{engle.2002} and \cite{tse_tsui.2002} which allows for time variation in the correlation of assets returns. In the last few years many models were also been proposed in order to improve the flexibility of the original DCC, see, e.g., \cite{bauwens_etal.2006} for a complete surveys of these recent developments. The success of the DCC model is due to its two stage estimation procedure which mainly relies on the multivariate Gaussian assumption. However, when the Gaussian assumption is violated, the separability of the log likelihood does not hold anymore and the only viable alternative in order to remain with a feasible model is to adopt a Quasi--MLE estimation approach, see, e.g., \cite{engle_sheppard.2001}. Despite its enormous popularity, only few works have been focused on the inclusion of exogenous information in a DCC framework in order to let the covariance co--movements to depend on covariates. Among those, \cite{vargas.2008} firstly introduced exogenous covariates into the Asymmetric Generalised DCC (AGDCC) model of \cite{cappiello_etal.2006}, while \cite{chou_liao.2008} follow a similar approach on a simpler DCC model. However, both the empirical applications of the aforementioned papers, only consider a bivariate setting under the Gaussian assumption for the innovation terms. Besides the tractability of the resulting model, one of the main difficulties of including exogenous covariates into a DCC model, is to guarantee the positiveness of the variance--covariance matrix at each point in time. A common solution relies on the Constrained Maximum Likelihood (CML) or Penalised Maximum Likelihood (PML) estimation approaches. In the context of bivariate time--varying copulas, only \cite{delirasalvatierra_patton.2015} consider the inclusion of exogenous information to model the correlation matrix dynamic. However, \cite{delirasalvatierra_patton.2015} do not consider a DCC--type dynamic evolution of the copula parameters, but, they instead rely on the GAS framework of \cite{creal_etal.2013} and \cite{harvey.2013}. Moreover, their model cannot be used in a fully multivariate setting when considering elliptical copulas such as the Gaussian and the Student--t. In what follows, we introduce our Student--t dynamic Markov--Switching (MS) copula model where the copula dependence matrix evolves according to a MS--DCC model which also depends on exogenous covariates.\newline
\indent Let $\bu_t=\left(u_{1,t},u_{2,t},\dots,u_{N,t}\right)^\prime$ with $u_{j,t}=F_{\mathcal{AST}}\left(y_{i,t};\btheta_{i,t}\right)$ be the Probability Integral Transformation (PIT) of $y_{i,t}$ according to the conditional AST distribution $F_{\mathcal{AST}}\left(\cdot\right)$ with parameters $\btheta_{i,t}=\left(\mu_{i,t},\sigma_{i,t},\gamma_{i,t},\nu_{i,t}\right)^\prime$, and assume that
\begin{equation}
\bu_t\mid S_t=s\sim\mathcal{C}_T\left(\bu_t;\bR_t^s,\nu_\sc^s\right),
\end{equation}
for $t=1,2,\dots,T$, where $\bR_t^s$ is the time--varying, regime dependent correlation matrix at time $t$ of the pseudo--observation $\bu_t$, and $\nu_\sc^s$ is the degree of freedom parameter that is also subject to the realisation of the first order Markov chain. According to the DCC dynamics the conditional correlation matrix $\bR_t^s$ can be decomposed in the following way
\begin{align}
\bR_t^s={\bD_t^s}^{-1}\bC_t^s{\bD_t^s}^{-1},
\end{align}
where $\bD_t^{s}$ is a diagonal matrix containing the square root of the diagonal elements of $\bC_t^s$. We further assume that the latent states $s=1,2,\dots,L$ are driven by a Markov process $S_t$, for $t=1,2,\dots,T$ defined on the discrete space $\Omega=\left\{1,2,\dots,L\right\}$ with transition probability matrix $\bQ=\left\{q_{l,k}\right\}$, where $q_{l,k}=\mathbb{P}\left(S_t=k\mid S_{t-1}=l\right)$, $\forall l,k\in\Omega$ is the probability that state $k$ is visited at time $t$ given that at time $t-1$ the chain was in state $l$, and initial probabilities vector $\bdelta=\left(\delta_1,\delta_2,\dots,\delta_L\right)^\prime$, $\delta_l=\mathbb{P}\left(S_1=l\right)$, i.e., the probability of being in state $l=\left\{1,2,\dots,L\right\}$ at time 1. For an up to date review of HMMs, see, e.g., \cite{cappe_etal.2005}, \cite{zucchini_macdonald.2009} and Dymarski \citeyearpar{dymarski.2011}. The following dynamics is imposed on the state dependent variance--covariance matrix
\begin{align}
\bC_{t+1}^s&= \left(\bar{\bC}-\bA^s\bar{\bC}{\bA^{s}}^\prime-\bB^s\bar{\bC}{\bB^{s}}^\prime-\bK{\bxi^{s}}^\prime\bar{\bX}\right)
+\bA^s\bXi_t{\bA^{s}}^\prime+\bB^s\bC_t^s{\bB^{s}}^\prime+{\bK\bxi^{s}}^\prime\bX_t,
\label{eq:corr_dynamic}
\end{align}
for $s=1,2,\dots,L$, where $\bA^s$ and $\bB^s$ are diagonal matrices with elements $\bA^s=\left\{\sqrt{\alpha_i^s}\right\}$, $\bB^s=\left\{\sqrt{\beta_i^s}\right\}$, for $i=1,2,\dots,N$, $s=1,2,\dots,L$, $\bK$ is a $N\times N$ matrix of ones, $\bxi^s$ is a $p\times 1$ vector containing the coefficients associated to the exogenous variables $\bX_t$. Here, $\bar\bC$ denotes the empirical variance--covariance matrix of the pseudo--observations $\bu_t$ and $\bar\bX$ is a $p\times 1$ vector containing the empirical mean of the regressors, which are introduced in equation \eqref{eq:corr_dynamic} in order to enforce the mean--reversion of the process. Moreover, in order to guarantee the stationarity of the process defined in equation \eqref{eq:corr_dynamic}, we impose $\alpha_i^s + \beta_i^s<1$ for all $i=1,2,\dots,N$ and $s=1,2,\dots,L$. Moreover, we need to ensure that $\bC_{t}^s$ is positive defined matrix for each $t=1,2,\dots,T$ and across all the possible states of nature $s=1,2,\dots,L$.
The only unspecified quantity in equation \eqref{eq:corr_dynamic}, is the forcing variable $\bXi_t$ which is set to the moving average variance--covariance matrix of length $m$
\begin{equation}
\bXi_t = m^{-1}\sum_{j=0}^m \bu_{t-m+j}\bu_{t-m+j}^\prime - \bar\bu_{m,j}\bar\bu_{m,j}^\prime,
\end{equation}
where $\bar\bu_{m,j} = m^{-1}\sum_{j=0}^m \bu_{t-m+j}$ represents the vector containing the simple mean of the pseudo observations across the period $\left\{t-m;t\right\}$. The proposed model is similar those of \cite{vargas.2008} and \cite{chou_liao.2008} but differs for the chosen forcing variable $\bXi_t$ which updates the conditional variance--covariance matrix. The proposed updating scheme makes use of a
rolling window variance--covariance matrix as forcing variable by which we can get a smooth time evolution of the correlation.
As regards this aspect, the proposed correlation dynamics generalises the approach of \cite{patton.2006b} and \cite{jondeau_rockinger.2006b} in a fully multivariate setting.\newline
\indent More interesting, the model here proposed, allows for the dynamic of the variance--covariance matrix to be also subject to the realisation of a first order Markov process. Using a different approach, \cite{bernardi_catania.2015a} show that including this regime dependent behaviour into the variance--covariance dynamics, really helps in describing the time--varying dependence of equity indexes. Indeed, ad documented, for example, by \cite{pelletier.2006}, \cite{guidolin_timmermann.2008}, \cite{ang_bekaert.2002}, \cite{bernardi_catania.2015a}, there is strong evidence of the presence of different regimes affecting the dependence structure of equity assets. Obviously, the model proposed in equation \eqref{eq:corr_dynamic} can be easily generalised to account for the leverage effect in a similar way as for the AGDCC of \cite{cappiello_etal.2006}
\begin{align}
\bC_{t+1}^s =&\left(\bar\bC-\bA^s\bar\bC{\bA^{s}}^\prime-\bB^s\bar\bC{\bB^{s}}^\prime-\bGamma^s\mathrm{\bar\bN}^s{\bGamma^{s}}^\prime-\bK{\bxi^{s}}^\prime\bar\bX \right)\nonumber\\
&\qquad\qquad\qquad+\bA^s\bXi_t{\bA^{s}}^\prime+\bB^s\bC_t^s{\bB^{s}}^\prime+\bGamma^s\boldsymbol\eta_t^s{\boldsymbol\eta_t^s}^\prime{\bGamma^{s}}^\prime+\bK{\bxi^{s}}^\prime\bX_t,
\label{eq:Gdynamic}
\end{align}
where $\bGamma^s=\left\{\sqrt{\gamma_i^s}\right\}$, $i=1,2,\dots,N$ is a diagonal matrix, and $\boldsymbol\eta_t^s$ is a vector whose $i$--th component is given by $\bbone\left(x_{i,t}^s<0\right)$ where $x_{i,t}^s=\mathcal{T}_{\nu_\sc^s}^{-1}\left(u_{it}\right)$ for $i=1,2,\dots,N$, and $\mathcal{T}_{\nu_\sc^s}^{-1}\left(\cdot\right)$ denotes the inverse cumulative density function (cdf) of a standardised Student--t distribution with $\nu_\sc^s$ degree of freedom, and $\bbone\left(\cdot\right)$ represents the indicator function.
In equation \eqref{eq:Gdynamic}, $\mathrm{\bar\bN}^s$ is the unconditional empirical average of the matrices $\boldsymbol\eta_t^s{\boldsymbol\eta_t^s}^\prime$, $t=1,2,\dots,T$. Several constraints must be imposed in order to avoid explosive patterns in the conditional variance--covariance matrix dynamic in equation \eqref{eq:Gdynamic}, see, e.g., \cite{cappiello_etal.2006}. Since in our empirical investigation we do not find any substantial improvement in favour of model \eqref{eq:Gdynamic} with respect to model \eqref{eq:corr_dynamic}, throughout the paper we will continue to refer to the specification without leverage effect. Moreover, since the number of parameters in the model in described in equation \eqref{eq:corr_dynamic} grows linearly with respect to the number of asset $N$, sometimes it would be convenient to set $\alpha_1^s = \alpha_2^s = \dots = \alpha_N^s=\alpha^s$ and $\beta_1^s = \beta_2^s = \dots = \beta_N^s=\beta^s$. We name this constrained specification \qmo simple\qmc, while we refer to the aforementioned general case with the name \qmo generalised\qmc.\newline
\indent To conclude the model specification, we report the multivariate density function of the random vector $\by_t=\left(y_{1,t},y_{2,t},\dots,y_{N,t}\right)^\prime$ which is given by
\begin{align}
\by_{t+1}\mid\by_{t}\sim h\left(\by_{t+1};\btheta_{t+1},\bX_{t},\bDelta_{t+1}^s, s=1,\dots,L\right),
\label{eq:multivariate_density}
\end{align}
where $\btheta_t=\left(\btheta_{1,t}^\prime,\btheta_{2,t}^\prime,\dots,\btheta_{N,t}^\prime\right)^\prime$, and
\begin{align}
h\left(\by_{y+1};\cdot\right)=&\prod_{i=1}^N f_{\mathcal{AST}}\left(y_{i,t+1};\btheta_{i,t+1}\right)
\sum_{s=1}^L\pi_{t+1\vert t}^{(s)} c_T\left(F_{\mathcal{AST}}\left(\by_{t+1};\btheta_{i,t+1}\right),\bDelta_{t+1}^s,\bX_t\right)\nonumber,
\end{align}
with $\bDelta_{t+1}^s=\left(\bR_{t+1}^s,\nu_{\sc}^s\right)$ and the mixing weight $\pi_{t+1\vert t}^{(s)}$ are
\begin{equation}
\pi_{t+1\vert T}^{(s)}=\sum_{m=1}^L {q}_{m,s}\mathbb{P}\left(S_{t+1}=m\mid\by_{1:t},\bX_{1:t}\right),\nonumber
\end{equation}
for $s=1,2,\dots,L$, and $\mathbb{P}\left(S_{t+1}=m\mid\br_{1:t},\bX_{1:t}\right)$ can be evaluated using the well known FFBS algorithm detailed in \cite{fruhwirth_schnatter.2006}. Here, the $i$--th marginal density and probability functions are represented by $f_\mathcal{AST}\left(\cdot\right)$ and $F_\mathcal{AST}\left(\cdot\right)$ while the copula probability density is denoted by $c_T\left(\cdot\right)$.\newline
\indent Concerning the estimation of the proposed model, unfortunately, given the presence of the exogenous covariates, we cannot rely on the usual results for DCC models such as, for examples, those of \cite{engle.2002} and \cite{cappiello_etal.2006}, while, the Constrained ML (CML) approach used by \cite{vargas.2008} and \cite{chou_liao.2008} in a similar context, requires the evaluation of highly nonlinear constraint during the maximisation procedure. We follow a different approach which consists on penalising the log--likelihood function (PML) in the second step of the IFM procedure, whenever the variance--covariance matrix is not positive defined. However, this solution is not optimal since it introduces strongly nonlinearities in the likelihood shape, and it may result in possible local optima solutions of the maximisation problem. Hence, good starting values are required.
In the second step of the IFM procedure, we made use of the Expectation--Maximisation algorithm of \cite{dempster_etal.1977} in order to deal with the markovian structure characterising the latent states. Further details about the employed estimation technique can be found in \cite{bernardi_catania.2015a}.
%
\section{Asset Allocation Strategy}
\label{sec:asset_allocation}
%
\noindent In this section, we tailor the \qmo distribution timing\qmcsp approach to the portfolio optimisation problem of \cite{jondeau_rockinger.2012} to the dynamic MS copula approach introduced in the previous Section. Specifically, we consider a rational investor having full information about the whole distribution of his/her future wealth and builds a dynamic asset allocation strategy by maximising his/her expected utility. It is worth noting that, higher moments play a fundamental role during the utility maximisation process.
Formally, let $y_{p,t+1} = \sum_{i=1}^N\lambda_{i,t\vert t+1} y_{i,t+1}$ be the future portfolio return, where $\lambda_{i,t\vert t+1}$ denotes the portion of today wealth that the investor is willing to allocate in the $i$--th risky asset, with $\sum_{i=1}^N \lambda_{i,t\vert t+1}=1$, then $W_{t+1} = 1 + y_{p,t+1}$ represents the future portfolio wealth at time $t+1$\footnote{We will always consider an initial wealth of 1\$. This choice will not affect the portfolio decision given our assumption on the investor's utility function.}. The vectors, $\by_{t+1} = \left(y_{1,t+1},y_{2,t+1},\dots,y_{N,t+1}\right)^\prime$ and $\blambda_{t\vert t+1}=\left(\lambda_{1,t\vert t+1},\lambda_{2,t\vert t+1},\dots,\lambda_{N,t\vert t+1}\right)^\prime$ contain all the assets in which the investor can open a speculative position at time $t$, which is subsequently held for the period $\left[t,t+1\right)$ and the portfolio weights, respectively. The portfolio optimisation problem can then be written as follows
\begin{equation}
\arg\max_{\left\{\blambda_{t\vert t+1}\right\}}\mathbb{E}_t\left[\mathcal{U}\left(1 + \blambda_{t\vert t+1}^\prime \by_{t+1}\right)\right],\nonumber
\label{eq:maximisation}
\end{equation}
where $\mathcal{U}\left(W_{t+1}\right)$ denotes the investor utility function and the expectation must be taken with respect to the future portfolio return distribution. Clearly, this problem does not have a closed--form solution given that $\by_{t+1}$ is conditionally distributed according to equation \eqref{eq:multivariate_density}. Moreover, when $N$ is large, numerical techniques such as, for example, Monte Carlo integration and quadrature rules become infeasible. Nevertheless, the maximisation problem can be easily solved by taking a Taylor expansion of the utility function around the wealth at time $t$, $W_t$, i.e., by considering
\begin{equation}
\mathcal{U}\left(W_{t+1}\right) = \sum_{k=0}^\infty\frac{\mathcal{U}^{\left(k\right)}\left(W_t\right)}{k!}\left(W_{t+1} - W_t\right)^k,
\label{eq:taylor_expansion}
\end{equation}
where $\mathcal{U}^{\left(k\right)}$ denotes the $k$--th derivative of the utility function with respect to its argument. Taking the expectation at time $t$ of both sides of equation \eqref{eq:taylor_expansion} we get
\begin{equation}
\mathbb{E}_t\left[\mathcal{U}\left(W_{t+1}\right)\right] =\sum_{k=0}^\infty\frac{\mathcal{U}^{\left(k\right)}\left(W_t\right)}{k!}m_{p,t+1}^{\left(k\right)},
\label{eq:taylor_expansion_portfolio}
\end{equation}
where $m_{p,t+1}^{\left(k\right)}=\mathbb{E}\left[r_{p,t+1}^k\right]=\mathbb{E}\left[\left(W_{t+1}-W_t\right)^k\right]$ denotes the $k$--th non central moment of the portfolio return distribution at time $t+1$. If we consider a fourth--order Taylor expansion, i.e., we truncate equation \eqref{eq:taylor_expansion_portfolio} at $\bar{k}=4$, the first four central moments of the portfolio distribution are immediately available as a function of the corresponding non--central moments and the portfolio weights. More precisely, the centred moments of the portfolio returns, at time $t+1$, are
\begin{align}
\mu_{p,t+1} &= \blambda_{t\vert t+1}^\prime M_{1,t+1}\nonumber\\
\sigma_{p,t+1}^2 &= \blambda_{t\vert t+1}^\prime M_{2,t+1}\blambda_{t\vert t+1}\nonumber\\
s_{p,t+1}^3 &= \blambda_{t\vert t+1}^\prime M_{3,t+1}\left(\blambda_{t\vert t+1}\otimes \blambda_{t\vert t+1}\right)\nonumber\\
k_{p,t+1}^4 &= \blambda_{t\vert t+1}^\prime M_{4,t+1}\left(\blambda_{t\vert t+1}\otimes \blambda_{t\vert t+1}\otimes \blambda_{t\vert t+1}\right),\nonumber
\end{align}
where $\otimes$ stands for the Kronecker product, and the corresponding non--centred moments required by equation \eqref{eq:taylor_expansion_portfolio} can be obtained as
\begin{align}
m_{p,t+1}^{\left(1\right)} &= \mu_{p,t+1} \nonumber\\
m_{p,t+1}^{\left(2\right)} &= \sigma_{p,t+1}^2 + \mu_{p,t+1}^2 \nonumber\\
m_{p,t+1}^{\left(3\right)} &= s_{p,t+1}^3  + 3\sigma_{p,t+1}^2\mu_{p,t+1} + \mu_{p,t+1}^3\nonumber\\
m_{p,t+1}^{\left(4\right)} &= k_{p,t+1}^4 + 4s_{p,t+1}^3\mu_{p,t+1} + 6\sigma_{p,t+1}^2\mu_{p,t+1}^2 + \mu_{p,t+1}^4.\nonumber
\end{align}
Note that $M_{1,t+1}$, $M_{2,t+1}$, $M_{3,t+1}$, $M_{4,t+1}$ are the mean, the variance--covariance matrix, the co--skewness and co--kurtosis of the conditional joint density function in equation \eqref{eq:multivariate_density}, respectively. Here, the original $\left(N,N,N\right)$ and $\left(N,N,N,N\right)$ matrices of co--skewness and co--kurtosis were transformed into the $\left(N,N^2\right)$ and  $\left(N,N^3\right)$ matrices, respectively, as suggested by \cite{deathayde_flores.2004} by slicing each $\left(N,N\right)$ layer and pasting them, in the same order, sideways.\newline
\indent Having defined the general portfolio optimisation framework, the only quantity that remains unspecified is the specific utility function the rational investor maximises. We assume that our agent makes his/her investment choices according to a Constant Relative Risk Aversion (CRRA) utility function defined as
\begin{equation}
%
\mathcal{U}_{t+1}\left(W_{t+1}\right) = \begin{cases} \frac{W_{t+1}^{1-\upsilon}}{1-\upsilon}, & \mbox{if } \upsilon>1 \\ \log\left(W_{t+1}\right), & \mbox{if } \upsilon=1, \end{cases}
\label{eq:utility}
\end{equation}
where $\upsilon$ is the coefficient of relative risk aversion, which is assumed to be constant with respect to the wealth $W_{t+1}$. For CRRA utility function, equation \eqref{eq:taylor_expansion_portfolio} truncated at the fourth order becomes
\begin{align}
\mathbb{E}_t\left[\mathcal{U}\left(W_{t+1}\right)\right]&\approx\frac{1}{1-\upsilon} + m_{p,t+1}^{\left(1\right)} - \frac{\upsilon}{2} m_{p,t+1}^{\left(2\right)}
+\frac{\upsilon\left(\upsilon + 1\right)}{6}  m_{p,t+1}^{\left(3\right)} -
\frac{\upsilon\left(\upsilon + 1\right)\left(\upsilon + 2\right)}{24} m_{p,t+1}^{\left(4\right)},
\label{eq:utility_final}
\end{align}
where the approximation order is $O\left(5\right)$. The closed--form expression for the expected utility makes the maximisation of equation \eqref{eq:utility_final} with respect to the portfolio weights, straightforward. However, unfortunately, there is no closed--form solution for the moments of order two to four of the multivariate joint density function and we need to resort to numerical procedures. \ref{sec:appendix_Moments} describes the method used to approximate the moments of the predictive joint density.
%
\section{Empirical study}
\label{sec:empirical_study}
%
\noindent In this section we apply the model and the portfolio optimisation methodology introduced in previous sections to deliver a set of optimal portfolio weights. To this end, we consider the same data set of \cite{jondeau_rockinger.2006a} which consists of five of the major international equity indexes, namely the Standard \& Poors 500 (SPX), the Nikkei 225 (N225), the FTSE 100 (FTSE), the DAX30 (GDAXI), and the CAC40 (FCHI). At each time $t$ over the out of sample period, the rational investor holding the portfolio is expected to choose the optimal allocation of the wealth by maximising his/her expected utility using the argument presented in Section \ref{sec:asset_allocation}. The illustration of the empirical results firstly focus on the marginal distribution, then deals with the multivariate density estimation and forecast and concludes with the presentation of the optimal portfolio allocations.
%
\subsection{Data}
\label{sec:data}
\noindent The data set consists of 1449 weekly returns spanning the period from the 8--th January, 1988, to the 9--th October, 2015. Starting from this series, the last 449 observations spanning the period from the 9--th March, 2007 to the end of the sample, are used to perform the out of sample analysis and to backtest the portfolio strategy, while the fist 1000 observation are used to fit the model and assess its in--sample performance. Table \ref{tab:Index_data_summary_stat} reports descriptive statistics of the indexes returns over the in--sample and out--of--sample periods. As expected, we find strong evidence of departure from normality, since all the considered series appear to be leptokurtic and skewed. Moreover, the \cite{jarque_bera.1980} statistic test strongly rejects the null hypothesis of normality for all the considered series at the 1\% confidence level. It is worth noting that, the departure from normality, appears to be stronger during the out--of--sample period. As discussed for example by \cite{shiller_2012}, this empirical evidence can be considered as an effect of the recent GFC of 2007--2008 that affected the overall economy. Table \ref{tab:Dependence} reports the linear correlation and the Kendall $\tau$ unconditional estimates. All the coefficients seem to be substantially different between the two subperiods, suggesting a time varying behaviour of the dependence structure that characterise our series. To further investigate this aspect, we perform the LMC test for constant versus time--varying conditional correlation of \cite{tse.2000} on the returns belonging to the whole period. In our case the LMC test is distributed according to a $\chi_{10}^2$ with a critical value of about 23.21 at the 1\% confidence level, while the test statistic is about 997 which strongly adverses the null hypothesis of time invariant correlation. Moreover, the lower triangular parts of Table \ref{tab:Dependence} also gives an insights about the changes in the tail of the joint probability distribution. These findings strongly support for the Student--t copula DCC model with exogenous regressors introduced in the previous sections \ref{sec:marginal_model}--\ref{sec:multivariate_model}, being able to describe the evolution of the whole joint distribution of assets returns.\newline
\indent As previously discussed, one of the main feature of the proposed model, concerns its ability to easily incorporate exogenous information in the state dependent correlation dynamics. Consequently, to control for the general economic conditions, we use observations of the following macroeconomic regressors as suggested by \cite{adrian_brunnermeier.2011}, \cite{chao_etal.2012} and \cite{bernardi_etal.2015}:
\begin{enumerate}
\item[(I)] the weekly change in the three--month Treasury Bill rate (3MTB);
\item[(II)] the weekly change in the slope of the yield curve (TERMSPR), measured by the difference of the 10-year Treasury rate and the 3--month Treasury Bill rate;
\item[(III)] the weekly change in the credit spread (CREDSPR) between 10--year BAA rated bonds and the 10-year Treasury rate.
\end{enumerate}
Historical data for all the considered exogenous regressors are freely available on the Federal Reserve Bank of St. Louis (FRED) web site.
%
\subsection{Parameters estimation and goodness--of--fit results}
\label{sec:parameters_est}
%
\noindent The empirical experiment is conducted as follows. First, we estimate the marginals and the MS copula DCC models over the in--sample period, and then we make a one--step ahead rolling forecast of the multivariate joint density of the returns for the entire out of sample period. During the rolling forecast exercise, the rational investor takes an investment decision in terms of wealth allocation, at each point in time $t+s$ conditional to the information set available at time $t+s-1$, for $s=1,2,\dots,448$. For both the marginal and copula models we re--estimate the model parameters each 24 observations, corresponding to six months using a fixed moving window.\newline
\indent Table \ref{tab:Marginal_estimates} reports the estimated coefficients for each GAS--AST marginal model. We find the scores coefficients associated with the scale $\left(\sigma\right)$ and shape $\left(\nu\right)$ parameters strongly significant indicating that our marginal model is able to capture the changes in the volatility patterns as well as in the tails of the conditional marginal distributions. On the contrary, the scores coefficients associated with the location $\left(\mu\right)$ and the skewness $\left(\gamma\right)$ parameters of the conditional AST distribution are not statistically different from zero. Moreover, we find that the shape and scale parameters of the AST distribution display high persistence, while the location and skewness parameters do not display such behaviour. The coefficients related to the scaled scores have quite similar magnitude across all the assets, indicating that the GAS updating mechanism adapts quite well to the different characteristics of the returns.\newline
\indent Before moving to the copula specification, we asses the goodness of fit of the marginal models. In particular, we want to test if the PITs of the estimated marginal densities are independently and identically distributed uniformly on the unit interval. To this end, we perform the same test employed by \cite{vlaar_palm.1993}, \cite{jondeau_rockinger.2006b}, and \cite{tay_etal.1998}. The test of the iid Uniform$\left(0,1\right)$ assumption consists of two parts.
The first part concerns the independent assumption, and it tests if all the conditional moments of the data, up to the fourth one, have been accounted for by the model, while the second part checks if the AST assumption is reliable by testing if the PITs are Uniform over the inverval $\left(0,1\right)$. In order to test if the PITs are independently and identically distributed, we define $\hat u_{i,t}=F_\mathcal{AST}\left(y_{i,t},\hat\btheta_{i,t}\right)$ as the PIT of return $i$ at time $t$ and $\bar{u}_i = T^{-1}\sum_{t=1}^T \hat u_{i,t}$ as the empirical average of the $i$--th PIT series, then we examine the serial correlation of $\left(\hat u_{i,t} - \bar{u}_i\right)^k$ for $k=1,2,\dots,4$ by regressing each $\left(\hat u_{i,t} - \bar{u}_i\right)$ for $i=1,2,\dots,N$ on its own lags up to the order 20.
The hypothesis of no serial correlation can be tested using a Lagrange multiplier test defined by the statistics $\left(T-20\right)R^2$ where $R^2$ is the coefficient of determination of the regression. This test is distributed according to a $\chi^2\left(20\right)$ with a critic value of about 31.4 at the 5\% confidence level. The first four columns of Table \ref{tab:Uniform_test}, named ${\rm DGT}-{\rm AR}^{\left(k\right)}$, report the test for $k=1,2,3,4$. Our results confirm that, for almost all the considered time series and all the moments of the univariate conditional distributions, the marginal models are able to effectively account for the dynamic behaviour of the series. For what concerns the Uniform assumption of the PITs, we again employ the test suggested by \cite{tay_etal.1998} which consists in splitting the empirical distribution of $\hat u_{i,t}$ into $G$ bins and test whether the empirical and the theoretical distributions significantly differ on each bin. More precisely, let us define $n_{q,i}$ as the number of $\hat u_{i,t}$ belonging to the bin $q$, it can be shown that
\begin{equation}
\xi_i=\frac{\sum_{q=1}^G\left(n_i-\mathbb{E} n_i\right)^2}{\mathbb{E} n_i} \sim \chi ^2\left(G-1\right),
\end{equation}
where $G$ is the number of bins. In our case, since we have estimated the model parameters, the asymptotic distribution of $\xi_i$ is bracketed between a $\chi ^2\left(G-1\right)$ and $\chi ^2\left(G-m-1\right)$ where $m$ is the number of estimated parameters.
In order to be consistent with the results reported by \cite{tay_etal.1998} and \cite{jondeau_rockinger.2006b}, we choose $G=20$ bins using a $\chi^2\left(19\right)$ distribution for $\xi_i,\quad i=1,2,\dots, N$, see also \cite{vlaar_palm.1993}. The test statistic of the Uniform test are reported in the last column of Table \ref{tab:Uniform_test} and are named $\mathrm{DGT-H}\left(20\right)$.
We can observe that, for each series, the PITs are uniformly distributed over the interval $(0,1)$ at a confidence level lower then $1\%$.
These findings definitely confirm the adequacy of our assumption on the innovation term and the conditional returns dynamic.\newline
%
\indent We now move to study the dependence structure of the series using the proposed copula model. As previously discussed, various parameterisations of the proposed model for the conditional dependence structure of assets returns can be employed. Given the available data set, several different levels of flexibility are possible. Model specification concerns the selection of either the number of states, or the \qmo simple\qmc, or \qmo generalised\qmc, specifications of the joint model discussed in Section \ref{sec:multivariate_model}, or the inclusion of the information coming from the exogenous data.
In particular, the choice of the number of regimes, in the HMM literature is an open question. As argued by \cite{lindsay.1983} and \cite{bohning.2000}, the number of states can be either estimated or tested. However given the large number of model combinations we are considering here, we decide to follow the most common practice of choosing the number of states according to the BIC, see e.g. \cite{cappe_etal.2005}, \cite{fruhwirth_schnatter.2006} and \cite{zucchini_macdonald.2009}. We chose the best model according to the BIC since we want to penalise more the models with a high number of parameters. Moreover, since the marginal models are invariant to the choice of the possible dependence specifications, we report the log--likelihood and the information criteria only for the copula specifications.
Table \ref{tab:Model_Choice} reports the log--likelihood evaluated at its optimum, the BIC and the AIC for different combinations of number of states, type of dynamic and inclusion of exogenous covariates. According to the BIC, the best model is the \qmc simple\qmcsp DCC dynamic specification with two regimes and the inclusion of the exogenous covariates. The BIC clearly select the model with the highest ability to represent different levels of the dependence structure affecting the data, and to react on the exogenous information depending on the state of the world. Table \ref{tab:CopulaCoef} reports the estimated coefficients for the best fitting model.
We can observe that both regimes are characterised by strong persistence in the dynamic of the conditional dependence, however the dynamics react differently to the new endogenous and exogenous information. More precisely, we can observe that in the second regime, the coefficient $\mathbf{A}^s$, $s=2$, which measures the reaction to new endogenous information, is much higher than the same coefficient in the other regime. This means that the second regime reacts more quickly, compared to the first one, when a change occurs in the market dependence structure. On the contrary, we observe a similar persistent behaviour of the correlation matrix within the two regimes. For what concerns the estimated state specific degree of freedom, we note that the second regime is characterised by a smaller coefficient indicating higher positive as well as negative tail dependence compared to the second one. These finding is also consistent with the empirical evidence reported by \cite{chollete_etal.2009}, \cite{bernardi_etal.2015} and \cite{bernardi_catania.2015a}, in related studies.
Looking at the estimated coefficients associated to the exogenous covariates, we can observe different behaviours of the conditional correlation matrix, depending on the regimes. In particular, we note that the returns dependence, conditional of being in the first regime, appears to react less to an absolute increase of the considered exogenous covariates. On the contrary, the dependence structure conditional of being in the second regime, displays a stronger reaction to the macroeconomic factors here considered.
%
%
The estimated transition probabilities of the unobserved Markov chain are $p_{11}=0.984$ and $p_{22}=0.987$, indicating quite strong persistence and an expected duration of the first and second states of about 61.5 and 75.9 weeks, respectively. Finally, we found that the dependence behaviour of the considered US market returns switched according to the 2000--2003 dot--com bubble and the recent 2008--2010 Global Financial Crisis.
%
\subsection{Portfolio allocation}
%
\noindent Now we consider the portfolio allocation problem. As stated in the Introduction, we focus on a rational investor who maximise his/her expected utility. Specifically, once the models parameters have been estimated over the in sample period, the investor forecasts the one--step ahead joint distribution of his/her assets return for each time $t$ of the out--of--sample period, and then he/she allocates the available wealth among those assets accordingly to the portfolio allocation procedure discussed in Section \ref{sec:asset_allocation}. We assume an initial wealth of 1\textdollar, an allocation strategy that allows for either long and short positions and a risk free rate equal to zero. The choice of the level of the initial wealth is otherwise irrelevant because it does not influence the risk aversion, and hence, the allocation strategy under the defined CRRA utility function reported in equation \eqref{eq:utility}. Moreover, in order to evaluate the influence of the level of risk aversion in the portfolio allocation strategy, we consider different levels of relative risk aversion, i.e., $\upsilon=3,7,10,20$. In order to asses the performance of the proposed model from a portfolio optimisation viewpoint, we run an horse race with five common alternatives:
\begin{itemize}
\item[-] the \qmo Equally Weighted portfolio\qmc, (EW) strategy, where the weights associated to each asset are kept fixed and equal to $1/N$ during the validation period;
\item[-] the \qmo Minimum Variance portfolio\qmc, (MV) strategy, where the weights associated to each asset are kept fixed and equal to those minimising the portfolio variance during the in--sample period;
\item[-] the \qmo DCC\qmcsp strategy, where the one--step ahead forecast of the conditional mean and the conditional variance--covariance matrix of the joint distribution of assets return are performed using the DCC$\left(1,1\right)$ model of \cite{engle.2002}, whose parameters are estimated by Quasi--ML, see, e.g., \cite{francq_zakoian.2011}. The weights of each asset are then chosen by maximising equation \eqref{eq:maximisation} truncated at the second order;
\item[-] the \qmo Naive Mean--Variance\qmc, (NMV) and \qmo Naive Higher--Moments\qmc, (NHM) strategies, where the optimal weights are chosen again by maximising equation \eqref{eq:maximisation}, as in the case of the \qmo DCC\qmcsp strategy, and the first four moments of the joint distribution of assets return are estimated using a rolling window approach.
\end{itemize}
Hereafter, we label the optimal investment strategy according to the model detailed in Section \ref{sec:model}, Flexible Dynamic Dependence Model (FDDM).\newline
\indent One easy way to compare different portfolios performance in terms of realised utility is to consider the so called \qmo management fee\qmcsp approach. The management fee approach relies on the definition of the amount of money, $\vartheta$, a rational investor is willing to pay, or gain, in order to switch from the portfolio that he/she is currently holding on a given period, to a different one, and it coincides with the solution of the the following equation
\begin{equation}
S^{-1}\sum_{t=F+1}^{F+S}\mathcal{U}\left(1+\blambda^{\mathcal{A}\prime}_{t\vert t+1}\by_{t+1}\right) = S^{-1}\sum_{t=F+1}^{F+S}\mathcal{U}\left(1+\blambda^{\mathcal{B}\prime}_{t\vert t+1}\by_{t+1}-\vartheta\right),
\label{eq:managementFee}
\end{equation}
where $F$ and $S$ are the length of the in--sample and out--of--sample periods, respectively, and, as before, $\blambda_{t\vert t+1}$ denotes the vector of amounts of wealth $\lambda_{j,t\vert t+1}$, the investor is allocating to each asset $j=1,2,\dots,N$ during the period $\left(t,t+1\right]$. From equation \eqref{eq:managementFee}, it is easy to see that if $\vartheta>0$, the investor is willing to pay a positive amount of money in order to switch from portfolio $\mathcal{A}$ to portfolio $\mathcal{B}$. On the contrary, if $\vartheta<0$, the investor is going to ask for a higher return from portfolio $\mathcal{B}$ in order to compensate the loss of utility he/she would experience for switching from $\mathcal{A}$ to $\mathcal{B}$. Finally, if $\vartheta=0$ the two portfolios give exactly the same utility to the investor, leaving he/she indifferent between the two options. Table \ref{tab:MF_mSR} reports the evaluated management fees for all the possible couples made by he FDDM strategy and a competitor, under different values of relative risk aversion coefficient $\upsilon=\left(3,5,10,20\right)$. More precisely, looking at equation \eqref{eq:managementFee}, we fix $\mathcal{B}$ equal to the FDDM strategy, and we let $\mathcal{A}$ to vary among the competing alternatives, i.e., the strategy a $\mathcal{A}$ is selected among the alternatives $\{\mathrm{EW},\mathrm{MV},\mathrm{DCC},\mathrm{NMV},\mathrm{NHM}\}$.
%
As a consequence, if the reported management fee is positive, then the optimal portfolio based on the proposed model is better compared to the specific competitor, otherwise the alternative strategy is preferred. Another useful indicator to compare portfolio strategies is the Sharpe ratio. However, since this indicator does not result in a measure of outperformance over alternative strategies with different levels of risk, we employ the so called modified Sharpe Ratio (mSR) introduced by \cite{graham_harvey.1997} which is defined as
\begin{equation}
\mathrm{mSR}\left(\mathcal{A},\mathcal{B}\right)=\frac{\sigma_{\mathcal{A}}}{\sigma_\mathcal{B}}\mu_\mathcal{B} - \mu_\mathcal{A},
\label{eq:modSharpeRatio}
\end{equation}
where it immediately follows that if $\mathrm{mSR}\left(\mathcal{A},\mathcal{B}\right)>0$, then model $\mathcal{B}$ is preferred to model $\mathcal{A}$ under a simple mean--variance preference ordering.\newline
\indent Table \ref{tab:MF_mSR} reports the management fee and the mSR for all the considered alternatives against the FDMM strategy. We note that, almost all the management fees and the Sharpe ratios are positive and significantly different from zero, indicating economic value in favour of the FDDM strategy. Indeed, concerning the evaluated management fees, we found that the FDDM strategy is almost always preferred by a rational investors with arbitrary level of risk aversion. The only exceptions are when the FDDM strategy is compared with the DCC strategy assuming an high degree of risk aversion $\left(\upsilon=3\right)$ and with the EW strategy assuming a low degree of risk aversion $\left(\upsilon=20\right)$. The results associated with the modified Sharpe ratio are in line with those reported by the management fees analysis. The only difference is that, when the degree of risk aversion is low, $\left(\upsilon=20\right)$, the statistical significance of the modified Sharpe ratio vanishes, indicating that there is no evidence of outperformance between the FDDM and the alternative strategies.
Table \ref{tab:avg_weights} reports the average portfolio weights over the whole out of sample period associated with each strategy. We note that the positions taken by the FDDM and DCC strategies are quite homogeneous in terms of which asset to buy and which asset to sell. Indeed, the N225 and the FCHI indexes are always shorted by the FDDM and DCC strategies. These finding suggest that the signal coming from the two strategies is similar, demonstrating that a truly dynamic model is important when investment positions are concerned. We also note that, the weights associated with the NMV, NHM, and MVP are very similar, independently from the assumed degree of risk aversion of the investor. Looking at the descriptive statistics of the indexes given in Table \ref{tab:Index_data_summary_stat} we can observe that the assets with the higher (in absolute value) wealth allocation, according to the FDDM strategy, are those who, ex--post, result in lower levels of kurtosis. On the contrary, the indexes with the lower exposition, report high kurtosis during the out--of--sample period. More interesting, looking again at Table \ref{tab:Index_data_summary_stat}, we note that, during the in--sample period, the levels of kurtosis among assets were quite homogeneous, which means that our model is able to effectively forecast the changes in the behaviour of the joint density. 
%
%
More precisely, on the one hand, the GAS updating mechanism is suited to effectively move the parameters through the direction of higher probability mass, and, on the other hand, the Markov switching dynamic copula model helps in capturing the evolution of the dependence structure during time and across states.
%
%
\section{Conclusion}
\label{sec:conclusion}
%
\noindent In this paper we propose a new flexible copula model being able to account for a wide variety of stylised facts affecting multivariate financial time series. Specifically, we allow the dependence structure affecting the considered financial time series to depend either on the realisation of a first order Markov chain and on exogenous covariates that may represent economic factors such as, interest rates, GDP growth rates, as well as financial indexes such as realised volatility indexes. The proposed solution is highly flexible, since the dependence dynamics account for either observable and unobservable latent factors.
%
Another interesting contribution concerns the marginal model specification. We develop a novel GAS filter \citep{creal_etal.2013, harvey.2013} based on the Asymmetric Student--t distribution of \cite{zhu_galbraith.2010} where all the location, scale and shape parameters evolves according to a first order transition dynamic based on the scaled score of the conditional distribution. In this way we allow for time variation in the first four moments of the conditional marginal distribution.
Parameters estimation is performed by exploiting the two step Inference Function for Margins procedure detailed in \cite{patton.2006a} for conditional copulas, where in the second step we made use of the Expectation--Maximisation algorithm of \cite{dempster_etal.1977}. Further details about the employed estimation technique can be found in \cite{bernardi_catania.2015a}.\newline
\indent The proposed model is applied to predict the evolution of the weekly returns of five major international stock indexes, previously considered by \cite{jondeau_rockinger.2006a}, over the period 1988--2015. In the empirical application, in sample and out of sample performances of the proposed model are deeply investigated. The in sample analysis, reveals statistical evidence for nonlinearities and highlights the relevance of the inclusion of exogenous covariates in explaining the dependence structure affecting the indexes returns. Furthermore, our analysis confirms that our model can be effectively used to describe and predict various aspects of the conditional joint distribution of equity returns, such as time varying means, variances, co--skewness and co--kurtosis. Out of sample model performances are evaluated by solving an optimisation problem where a rational investor with power utility function takes future investment positions according to the predictive density of assets returns. 
Specifically, we compare the optimal portfolio strategy implied by the proposed model with that of several static and dynamic alternatives. The portfolio optimisation strategy is based on the Taylor expansion of a CRRA utility function of the rational investor and involves the evaluation of the portfolio moments up to the fourth order. Since higher order moments of the joint distribution are not analytically known we resort to numerical integration. Our empirical results suggest that the proposed model outperforms competitors in terms of economic value.
%
%
%
\section*{Acknowledgments}
\noindent This research is supported by the Italian Ministry of Research PRIN 2013--2015, ``Multivariate Statistical Methods for Risk Assessment'' (MISURA), and by the ``Carlo Giannini Research Fellowship'', the ``Centro Interuniversitario di Econometria'' (CIdE) and ``UniCredit Foundation''. We would like to express our sincere thanks to all the participants to the Bozen -- Risk School, Bozen, September 2015,  for their constructive comments that greatly contributed to improving the final version of the paper. A special thank goes also to Juri Marcucci and to all the participants of the 2015 Workshop in Econometrics and Empirical Economics (WEEE) held at the SADiBA, Perugia 27--28 August 2015, organised by the CIdE and the Bank of Italy.
%
\appendix
%
\section{Evaluation of higher moments}
\label{sec:appendix_Moments}
\noindent In this appendix we discuss how to evaluate the quantities denoted by $M_{1,t}$, $M_{2,t}$, $M_{3,t}$, $M_{4,t}$, for $t=1,2,\dots,T$, when the number of asset $N$ is large. Clearly, the first moment $M_{1,t}$ is directly available once the marginal distributions are specified. In our context, given the AST marginal specification, the first moment can be found in \cite{zhu_galbraith.2010}.
Problems arise as a direct consequence of the copula specification, when we want to evaluate the remaining moments $M_{2,t}$, $M_{3,t}$, and $M_{4,t}$. Specifically, let $\by_t=\left(y_{1,t},y_{2,t},\dots,y_{N,t}\right)^\prime$ be the vector of assets returns at time $t$, we define
\begin{align}
\sigma_{ijt} &= \mathbb{E}\left[\left(y_{i,t} - \mu_{i,t}\right)\left(y_{j,t} - \mu_{j,t}\right)\right]\nonumber\\
s_{ijk} &=  \mathbb{E}\left[\left(y_{i,t} - \mu_i\right)\left(y_{j,t} - \mu_j\right) \left(y_{k,t} - \mu_{k,t}\right)\right]\nonumber\\
k_{ijkl} &=  \mathbb{E}\left[\left(y_{i,t} - \mu_{i,t}\right)\left(y_{j,t} - \mu_{j,t}\right) \left(y_{k,t} - \mu_{k,t}\right) \left(y_{l,t} - \mu_{l,t}\right)\right],\nonumber
\end{align}
where the expectation is taken with respect to the predictive distribution defined in equation \eqref{eq:multivariate_density}. A first solution is given by numerical integration, i.e., considering
\begin{equation}
\hat M_q = \int_{\mathbb{R}^N}M_q\left(\by_t\right) h\left(\by_t\mid\by_{1:t-1},\bX_{1:t-1}\right)d\by_t,\qquad q=2,3,4,
\label{eq:joint_moments}
\end{equation}
where $\by_{1:t-1}$ and $\bX_{1:t-1}$ denote observations up to time $t-1$. However we found this solution really imprecise and time consuming when N is moderately large. Another interesting solution is to rely on simulation methods, i.e., simulate $B$ draws from $h\left(\by_t\mid\by_{1:t-1},\bX_{1:t-1}\right)$ and then compute $\hat M_2$, $\hat M_3$, and $\hat M_4$ as the empirical counterpart of the corresponding theoretical moment defined in equation \eqref{eq:joint_moments}. This solution is easy to implement and reported good results even for large system. 
%
\section{Score vector of the AST distribution}
\label{sec:ScoreAndFisher}
%
\noindent In this appendix we report the score vector of the AST distribution which is needed in order to compute the GAS--AST dynamics. The score of the AST distribution at time $t$ evaluated at $y^0$ given the parameters vector $\vartheta=\left(\mu,\sigma,\gamma,\nu\right)$ is
\begin{equation}
\frac{\partial\log f_\mathcal{AST}\left(y^0;\vartheta\right)}{\partial \vartheta}=\left(\frac{\partial \log f_\mathcal{AST}\left(y^0;\vartheta\right)}{\partial \mu},\frac{\partial \log f_\mathcal{AST}\left(y^0;\vartheta\right)}{\partial \sigma},\frac{\partial \log f_\mathcal{AST}\left(y^0;\vartheta\right)}{\partial \gamma},\frac{\partial \log f_\mathcal{AST}\left(y^0;\vartheta\right)}{\partial \nu}\right)^\prime,\nonumber
\end{equation}
where
\begin{align}
  \frac{\partial \log f_\mathcal{AST}}{\partial \mu} =& \begin{cases}
  \frac{\nu + 1}{A_1\left(y^0\right)}\frac{y^0-\mu}{\nu\left(2\gamma\sigma K\left(y^0\right)\right)^2}&\quad\mathrm{if}\quad y^0\le\mu\\
  \frac{\nu + 1}{A_2\left(y^0\right)}\frac{y^0-\mu}{\nu\left(2\left(1-\gamma\right)\sigma K\left(y^0\right)\right)^2}&\quad\mathrm{if}\quad y^0>\mu\\
  \end{cases}\nonumber\\
  \frac{\partial \log f_\mathcal{AST}}{\partial \sigma} =& \begin{cases}
  -\frac{1}{\sigma} + \frac{\nu + 1}{A_1\left(y^0\right)\sigma^3\nu} \left[\frac{y^0-\mu}{\nu\left(2\gamma\sigma K\left(y^0\right)\right)^2}\right]^2 &\quad\mathrm{if}\quad y^0\le\mu\nonumber\\
  -\frac{1}{\sigma} + \frac{\nu + 1}{A_2\left(y^0\right)\sigma^3\nu} \left[\frac{y^0-\mu}{\nu\left(2\left(1-\gamma\right)\sigma K\left(y^0\right)\right)^2}\right]^2 &\quad\mathrm{if}\quad y^0>\mu\nonumber\\
  \end{cases}\nonumber\\
  \frac{\partial \log f_\mathcal{AST}}{\partial \gamma} =& \begin{cases}
  \frac{\nu + 1}{A_1\left(y^0\right)\gamma^3\nu^2} \frac{y^0-\mu}{\left(2\sigma K\left(y^0\right)\right)^2} &\quad\mathrm{if}\quad y^0\le\mu\\
  \frac{\nu + 1}{A_2\left(y^0\right)\left(1-\gamma\right)^3\nu^2} \frac{y^0-\mu}{\left(2\sigma K\left(y^0\right)\right)^2} &\quad\mathrm{if}\quad y^0>\mu\nonumber\\
  \end{cases}\\
  \frac{\partial \log f_\mathcal{AST}}{\partial \nu} =& \begin{cases}
  -\left[\frac{\log A_1\left(y^0\right)}{2} +   \frac{\nu+1}{2A_1\left(y^0\right)}\left(  -\frac{\left(y^0-\mu\right)^2}{\left(2\nu\gamma\sigma K\left(y^0\right)\right)^2 }  +  \frac{\left(y^0-\mu\right)^2}{K\left(y^0\right)^3\nu\left(2\gamma\sigma\right)^2} \left(-2\frac{\partial K\left(y^0\right)}{\partial y^0}\right)      \right)  \right] &\quad\mathrm{if}\quad y^0\le\mu\nonumber\\
  -\left[\frac{\log A_2\left(y^0\right)}{2} +   \frac{\nu+1}{2A_2\left(y^0\right)}\left(  -\frac{\left(y^0-\mu\right)^2}{\left(2\nu\left(1-\gamma\right)\sigma K\left(y^0\right)\right)^2 }  +  \frac{\left(y^0-\mu\right)^2}{K\left(y^0\right)^3\nu\left(2\left(1-\gamma\right)\sigma\right)^2} \left(-2\frac{\partial K\left(y^0\right)}{\partial y^0}\right)      \right)  \right] &\quad\mathrm{if}\quad y^0>\mu,\nonumber\\
  \end{cases}\nonumber
\end{align}
and $A_1\left(y^0\right) = 1 + \left[\frac{y^0-\mu}{2\gamma\sigma K\left(y^0\right)}\right]^2$, $A_2\left(y^0\right) = 1 + \left[\frac{y^0-\mu}{2\left(1-\gamma\right)\sigma K\left(y^0\right)}\right]^2$. The variance of the score of the AST distribution at time $t$ evaluated at $y^0$ is
%
\section{Tables}
\label{sec:appendix_A}
%
\begin{table}[!ht]
\captionsetup{font={small}, labelfont=sc}
\begin{center}
\begin{small}
\smallskip
\begin{tabular}{lccccccccc}\\
\toprule
Name & Min & Max & Mean & Std. Dev. & Skewness & Kurtosis & 1\% Str. Lev. & JB \\
\hline
\multicolumn{9}{l}{\textit{In sample, from 08/01/1988 to 02/03/2007}}\\
SPX & -12.33 & 7.49 & 0.17 & 2.06 & -0.47 & 5.98 & -5.12 & 406.29 \\
N225 & -12.5 & 11.05 & -0.02 & 2.83 & -0.13 & 4.41 & -7.13 & 85.17 \\
FTSE & -8.86 & 10.07 & 0.13 & 2.06 & -0.06 & 4.67 & -5.58 & 116.46 \\
GDAXI & -13.92 & 12.89 & 0.19 & 2.89 & -0.32 & 4.93 & -7.68 & 172.68 \\
FCHI & -12.13 & 11.03 & 0.17 & 2.7 & -0.14 & 4.15 & -6.73 & 58.14 \\
\hline
Name & Min & Max & Mean & Std. Dev. & Skewness & Kurtosis & 1\% Str. Lev. & JB \\
\hline
\multicolumn{9}{l}{\textit{Out of sample, from 09/03/2008 to 09/10/2015}}\\
SPX & -20.08 & 11.36 & 0.08 & 2.71 & -0.93 & 11.07 & -7.21 & 1283.12 \\
N225 & -27.88 & 11.45 & 0.02 & 3.32 & -1.54 & 14.18 & -8.01 & 2517.6 \\
FTSE & -23.63 & 12.58 & 0.01 & 2.78 & -1.38 & 16.33 & -7.77 & 3468.48 \\
GDAXI & -24.35 & 14.94 & 0.09 & 3.38 & -1.02 & 10.59 & -9.31 & 1155.59 \\
FCHI & -25.05 & 12.43 & -0.03 & 3.33 & -1.19 & 10.69 & -8.88 & 1212.1 \\
\bottomrule
\end{tabular}
%
\caption{\footnotesize{Summary statistics of the five weekly equity indexes returns in percentage points, over the period starting form January, 8th 1988 to October, 9st 2015. The seventh column, denoted by ``1\% Str. Lev.'' is the 1\% empirical quantile of the returns distribution, while the eight column, denoted by ``JB'' is the value of the Jarque-Ber\'a test-statistics.}}
\label{tab:Index_data_summary_stat}
\end{small}
\end{center}
\end{table}
%
\begin{table}[!t]
\captionsetup{font={small}, labelfont=sc}
\begin{small}
\centering
\smallskip
\begin{tabular}{lccccccccccccccccccccc}\\
\toprule
\multicolumn{6}{l}{\textit{In sample, from 28/12/2001 to 26/01/2011}}\\
\hline
$\tau/\rho$ & SPX & N225 & FTSE & GDAXI & FCHI\\
SPX & -- & 0.33 & 0.62 & 0.6 & 0.59 \\
N225 & 0.23 & -- & 0.36 & 0.36 & 0.36 \\
FTSE & 0.42 & 0.25 & -- & 0.68 & 0.69 \\
GDAXI & 0.25 & 0.46 & 0.25 & -- & 0.79 \\
FCHI & 0.41 & 0.41 & 0.5 & 0.58 & -- \\
\hline
\multicolumn{6}{l}{\textit{Out of sample, from 27/01/2011 to 30/01/2015}}\\
\hline
$\tau/\rho$  & SPX & N225 & FTSE & GDAXI & FCHI \\
SPX & -- & 0.67 & 0.87 & 0.85 & 0.68 \\
N225 & 0.41 & -- & 0.84 & 0.68 & 0.89 \\
FTSE & 0.64 & 0.42 & -- & 0.67 & 0.92 \\
GDAXI & 0.61 & 0.42 & 0.66 & -- & 0.93 \\
FCHI & 0.64 & 0.44 & 0.71 & 0.77 & -- \\
\bottomrule
\end{tabular}
%
\caption{\footnotesize{Linear correlation and Kendal's $\tau$ coefficients calculated on the in sample and out of sample periods.}}
\label{tab:Dependence}
\end{small}
%
\end{table}
%
\begin{table}[!th]
\centering
\resizebox{1.0\columnwidth}{!}{%
\begin{tabular}{lccccccccccccc}
\toprule
Asset & $\omega_\mu$ & $\omega_\sigma$ & $\omega_\alpha$ & $\omega_\nu$ & $\alpha_\mu$ & $\alpha_\sigma$ & $\alpha_\alpha$ & $\alpha_\nu$ & $\beta_\mu$ & $\beta_\sigma$ & $\beta_\alpha$ & $\beta_\nu$ \\
\cmidrule(lr){1-1}\cmidrule(lr){2-13}
SPX & $0.0043^a$ & $-0.2327^b$ & $0.1816^c$ & $-0.7568^b$ & $0$ & $0.0769^a$ & $0.0134$ & $0.0287^a$ & $0.2368^c$ & $0.9712^a$ & $0.339$ & $0.6759^a$ \\
N225 & $0.0068$ & $-0.5661^a$ & $0.0767$ & $-0.4471^b$ & $0.0091$ & $0.0754^a$ & $0.006$ & $0.0241^a$ & $0$ & $0.9274^a$ & $0.7514^b$ & $0.825^a$ \\
FTSE & $0.0033^a$ & $-0.3253^a$ & $8e-04$ & $-0.0409^b$ & $0$ & $0.0576^a$ & $0.0023$ & $0.0133^a$ & $0.1305^a$ & $0.9598^a$ & $0.9947^a$ & $0.9803^a$ \\
GDAXI & $0.0077^a$ & $-0.4727^a$ & $0.2719^b$ & $-0.3581^a$ & $0$ & $0.0982^a$ & $0.0163$ & $0.0178^a$ & $0.1493^a$ & $0.9396^a$ & $0.2313$ & $0.8641^a$ \\
FCHI & $0.0069^a$ & $-0.2414^a$ & $0.3212^a$ & $-0.2703^b$ & $0$ & $0.0639^a$ & $0.0104$ & $0.0123^a$ & $0.1214$ & $0.9691^a$ & $0.0497$ & $0.8903^a$ \\
\bottomrule
\end{tabular}}
\caption{\footnotesize{Parameters estimate of the marginal GAS--AST models, for the in sample period form December, 28th 2001 to January, 26th 2011. The apexes \qmo a\qmc, \qmo b\qmcsp and \qmo c\qmc, denote the rejection of the null hypothesis of not significance of the corresponding parameter, at different confidence levels $1\%$, $5\%$ and $10\%$.}}
\label{tab:Marginal_estimates}
\end{table}
%
%
\begin{table}[!t]
\captionsetup{font={small}, labelfont=sc}
\begin{small}
\centering
\begin{tabular}{lcccccc}\\
\toprule
State &$\mathbf{A}^s$ & $\mathbf{B}^s$  & $\nu_\sc^s$    & (3MTB)$^s$ &(CREDSPR)$^s$  &(TERMSPR)$^s$ \\
\cmidrule(lr){1-1}\cmidrule(lr){2-7}
1 & $0.02250^a$ & $0.88913^a$ & $34.53073^a$ & $0.00005^a$ & $ 0.00032^a$ & $ -0.00005^a$\\
2 & $0.00001^a$ & $0.87940^a$ & $29.84355^a$ & $0.00073^a$ & $ 0.00430^a$ & $ -0.00069^a$\\
\bottomrule
\end{tabular}
%
\caption{\footnotesize{Parameters estimate of the DCC--Copula model with exogenous regressors, for the in sample period form December, 28th 2001 to January, 26th 2011. The apexes \qmo a\qmc, \qmo b\qmcsp and \qmo c\qmc, denote the rejection of the null hypothesis of not significance of the corresponding parameter, at different confidence levels $1\%$, $5\%$ and $10\%$.}}
\label{tab:CopulaCoef}
\end{small}
%
\end{table}
\begin{table}[!th]
\centering
\begin{tabular}{lccccc}
\hline
Asset & $\mathrm{DGT-AR}^{\left(1\right)}$ & $\mathrm{DGT-AR}^{\left(2\right)}$ & $\mathrm{DGT-AR}^{\left(3\right)}$ & $\mathrm{DGT-AR}^{\left(4\right)}$ & $\mathrm{DGT-H}\left(20\right)$ \\
\cmidrule(lr){1-1}\cmidrule(lr){2-6}
SPX & $34.54^b$ & $24.38$ & $37.65^a$ & $23.32$ & $13.12$ \\
N225 & $13.31$ & $18.33$ & $12.47$ & $19.46$ & $12.04$ \\
FTSE & $22.6$ & $22.01$ & $22.38$ & $23.39$ & $9.76$ \\
GDAXI & $13.71$ & $26.01$ & $16.21$ & $25.03$ & $21.6$ \\
FCHI & $23.99$ & $24.19$ & $15.91$ & $23.53$ & $12.92$ \\
\hline
\end{tabular}
\caption{\footnotesize{In sample Goodness--of--Fit test of \cite{tay_etal.1998}.  The apexes \qmo a\qmc, \qmo b\qmcsp and \qmo c\qmc, denote the rejection of the null hypothesis of not significance of the corresponding parameter, at different confidence levels $1\%$, $5\%$ and $10\%$. See also \cite{vlaar_palm.1993} and \cite{jondeau_rockinger.2006b}.}}
\label{tab:Uniform_test}
\end{table}
%
%
\begin{table}[!t]
\captionsetup{font={small}, labelfont=sc}
\begin{center}
\begin{small}
\smallskip
\begin{tabular}{lcccccccc}\\
\toprule
 Model & \multicolumn{4}{c}{Without exogenous regressors} & \multicolumn{4}{c}{With exogenous regressors}\\
\cmidrule(lr){2-5}\cmidrule(lr){6-9}
 & AIC & BIC & $\ell_T$ & $n_p$ & AIC & BIC & $\ell_T$ & $n_p$\\
\hline
\multicolumn{9}{l}{\textit{Simple}}\\
1& 	2715.48	& 2718.96& 	-1354.74& 	3& 	2714.64	& 2722.77& 	-1350.32& 	7 \\
2& 	2705.75	& 2715.04& 	-1344.88& 	8& 	2680.44& 	$\mathbf{2699.02}$& 	-1324.22& 	16 \\
3& 	2711.46	& 2728.88& 	-1340.73& 	15	& 2675.90	& 2707.25	& -1310.95	& 27 \\
\hline
\multicolumn{9}{l}{\textit{Generalised}}\\
1	& 2692.28& 	2705.05& 	-1335.14& 	11	& 2697.16	& 2714.58& 	-1333.58	& 15 \\
2	& 2703.26& 	2731.13& 	-1327.63& 	24& 	2694.42& 	2731.57	& -1315.21	& 32 \\
3	& 2726.22& 	2771.50& 	-1324.11& 	39& 	2713.46& 	2772.67	& -1305.73	& 51 \\
\bottomrule
\end{tabular}
%
\caption{\footnotesize{AIC, BIC, log--likelihood $\left(\ell_T\right)$ and number of parameters $\left(n_p\right)$, for the simple FDDM and the generalised FDMM model specifications with and without the inclusion of exogenous regressors. The first column denotes the number of hidden states.}}
\label{tab:Model_Choice}
\end{small}
\end{center}
%
\end{table}
%
%
\begin{table}[!th]
\centering
\begin{tabular}{lccccc}
\hline
Strategy & SPX & N225 & FTSE & GDAXI & FCHI \\
\cmidrule(lr){1-6}
\multicolumn{6}{c}{$\upsilon=3$}\\
\cmidrule(lr){1-1}\cmidrule(lr){2-6}
FDD & 0.31 & -0.55 & 0.64 & 2.8 & -2.2 \\
DCC & 0.21 & -0.81 & -0.2 & 3.65 & -1.86 \\
NMV & 1.1 & -0.61 & 0.07 & 0.29 & 0.15 \\
NHM & 1.07 & -0.59 & 0.09 & 0.28 & 0.16 \\
\cmidrule(lr){1-1}\cmidrule(lr){2-6}
\multicolumn{6}{c}{$\upsilon=7$}\\
\cmidrule(lr){1-1}\cmidrule(lr){2-6}
FDD & 0.39 & -0.12 & 0.64 & 1.22 & -1.12 \\
DCC & 0.35 & -0.23 & 0.36 & 1.59 & -1.07 \\
NMV & 0.73 & -0.15 & 0.3 & 0.06 & 0.06 \\
NHM & 0.7 & -0.13 & 0.31 & 0.05 & 0.07 \\
\cmidrule(lr){1-1}\cmidrule(lr){2-6}
\multicolumn{6}{c}{$\upsilon=10$}\\
\cmidrule(lr){1-1}\cmidrule(lr){2-6}
FDD & 0.41 & -0.02 & 0.63 & 0.86 & -0.87 \\
DCC & 0.38 & -0.1 & 0.48 & 1.13 & -0.89 \\
NMV & 0.65 & -0.04 & 0.35 & 0.01 & 0.04 \\
NHM & 0.6 & -0.03 & 0.37 & -0.01 & 0.06 \\
\cmidrule(lr){1-1}\cmidrule(lr){2-6}
\multicolumn{6}{c}{$\upsilon=20$}\\
\cmidrule(lr){1-1}\cmidrule(lr){2-6}
FDD & 0.44 & 0.09 & 0.62 & 0.43 & -0.58 \\
DCC & 0.42 & 0.05 & 0.61 & 0.61 & -0.68 \\
NMV & 0.55 & 0.07 & 0.41 & -0.05 & 0.02 \\
NHM & 0.49 & 0.1 & 0.44 & -0.07 & 0.04 \\
\cmidrule(lr){1-1}\cmidrule(lr){2-6}
MVP & 0.45 & 0.2 & 0.47 & -0.12 & 0 \\
1ON & 0.20 & 0.20 & 0.20 & 0.20 & 0.20\\
\hline
\end{tabular}
\caption{\footnotesize{Average portfolio weights over the out of sample period from from January 27, 2011 to January 30, 2015.}}
\label{tab:avg_weights}
\end{table}
%
\begin{table}[!t]
\captionsetup{font={small}, labelfont=sc}
\begin{center}
\begin{small}
\smallskip
\begin{tabular}{lcccccccc}
\toprule
& \multicolumn{4}{c}{Management Fee} & \multicolumn{4}{c}{Modified SR}\\
\cmidrule(lr){1-1}\cmidrule(lr){2-5}\cmidrule(lr){6-9}
Strategy & $\upsilon=3$ & $\upsilon=7$ & $\upsilon=10$ & $\upsilon=20$ & $\upsilon=3$ & $\upsilon=7$ & $\upsilon=10$ & $\upsilon=20$  \\
\cmidrule(lr){1-1}\cmidrule(lr){2-5}\cmidrule(lr){6-9}
EW & $1.12^a$ & $0.386^b$ & $0.134$ & $-0.362^a$ & $0.33^b$ & $0.139^c$ & $0.095^c$ & $0.04$  \\
DCC & $-0.337^c$ & $-0.041$ & $0.109^c$ & $0.477^a$ & $-0.016$ & $0.002$ & $0.005$ & $0.008$ \\
MV & $0.691^c$ & $0.338^c$ & $0.347^a$ & $0.554^a$ & $0.161^c$ & $0.111^c$ & $0.083^c$ & $0.037$  \\
NMV & $0.511$ & $0.261^c$ & $0.293^b$ & $0.526^a$ & $0.149^c$ & $0.09^c$ & $0.066$ & $0.027$ \\
NHM & $0.52^c$ & $0.27^c$ & $0.305^b$ & $0.543^a$ & $0.151^c$ & $0.093^c$ & $0.071$ & $0.033$ \\
\bottomrule
\end{tabular}
%
\caption{\footnotesize{Management fee that a rational investor is willing to pay for switching between the FDDM versus various alternatives and the modified Sharpe Ratio. The apexes \qmo a\qmc, \qmo b\qmcsp and \qmo c\qmc, denote the rejection of the null hypothesis of not significance of the corresponding parameter, at different confidence levels $1\%$, $5\%$ and $10\%$. P--values are obtained using block bootstrap techniques.}}
\label{tab:MF_mSR}
\end{small}
\end{center}
\end{table}
\clearpage

%
\bibliographystyle{apalike}

\end{document}